\documentclass[prd,superscriptaddress,twocolumn,preprintnumbers,nofootinbib,%
  amsmath,amssymb,longbibliography]{revtex4-1}
  
\usepackage[breaklinks,colorlinks=true]{hyperref}
\usepackage[T1]{fontenc}
\usepackage[utf8]{inputenc}
\usepackage{mathtools}
\usepackage{amsmath,amsthm,amssymb}
\usepackage{amsmath,bm}
\usepackage{bbold} 
\usepackage{dsfont}
\usepackage{lipsum}
\usepackage{calrsfs}
\DeclareMathAlphabet{\pazocal}{OMS}{zplm}{m}{n}

\usepackage{color}
\usepackage{hyperref}
\usepackage{bm}
\allowdisplaybreaks

\usepackage[dvipsnames]{xcolor}
\newcommand\myshade{85}
\colorlet{mylinkcolor}{BrickRed}
\colorlet{mycitecolor}{NavyBlue}
\colorlet{myurlcolor}{Aquamarine}
\hypersetup{
  linkcolor  = mylinkcolor!\myshade!black,
  citecolor  = mycitecolor!\myshade!black,
  urlcolor   = myurlcolor!\myshade!black,
  colorlinks = true,
}

\newcommand\subref[2]{\hyperref[#1]{\ref*{#1}#2}}

\begin{document}
\title{Dynamical noise can enhance high-order statistical structure in complex systems}

\author{Patricio Orio}
\email{Email: patricio.orio@uv.cl}
\affiliation{Centro Interdisciplinario de Neurociencia de Valparaíso, Universidad de Valparaíso, Chile}%2360103 Valparaíso, Chile}
\affiliation{Instituto de Neurociencia, Facultad de Ciencias, Universidad de Valparaíso, Chile}%2360102 Valparaíso, Chile}

\author{Pedro A.M. Mediano}
\affiliation{Department of Computing, Imperial College London, UK}
\affiliation{Department of Psychology, University of Cambridge, UK}

\author{Fernando E. Rosas}
\affiliation{Department of Informatics, University of Sussex, UK}
\affiliation{Centre for Psychedelic Research, Department of Brain Science, Imperial College London, UK}
\affiliation{Centre for Complexity Science, Imperial College London, UK}
\affiliation{Centre for Eudaimonia and Human Flourishing, University of Oxford, UK}

\newtheorem{definition}{Definition}
\newtheorem{theorem}{Theorem}
\newtheorem{lemma}{Lemma}
\newtheorem{proposition}{Proposition}
\newtheorem{corollary}{Corollary}
\newtheorem{example}{Example}
\newtheorem{remark}{Remark}

\begin{abstract}

\noindent
Recent research has provided a wealth of evidence highlighting the pivotal role
of high-order interdependencies in supporting the information-processing capabilities of distributed complex systems. 
These findings may suggest that high-order interdependencies constitute a powerful 
resource that is, however, challenging to harness and can be readily disrupted. 
In this paper we contest this perspective by demonstrating that high-order
interdependencies can not only exhibit robustness to stochastic perturbations, but can in fact be enhanced by them. 
Using elementary cellular automata as a general testbed, our results unveil the
capacity of dynamical noise to enhance the statistical regularities between
agents and, intriguingly, even alter the prevailing character of their
interdependencies. Furthermore, our results show that these effects are related
to the high-order structure of the local rules, which affect the system's
susceptibility to noise and characteristic times-scales. 
These results deepen our understanding of how high-order interdependencies may
spontaneously emerge within distributed systems interacting with stochastic
environments, thus providing an initial step towards elucidating their origin
and function in complex systems like the human brain.

\end{abstract}

\maketitle

\section{Introduction}

Complex systems are characterised by collective phenomena arising from the
interactions between relatively simple elements~\cite{waldrop1993complexity}.
Examples include the mesmerising synchronisation of fireflies and the
murmuration of starlings~\cite{strogatz2004sync}, the freezing of water and
other phase transitions~\cite{sole2011phase}, and the collective dynamics of
decision-making~\cite{rosas2017technological} or political
opinions~\cite{rajpal2019tangled}. Crucially, these collective phenomena often
cannot be explained from the behaviour of the system's parts in isolation. Thus,
the goal of complexity science is to identify common laws governing the
emergence of collective behaviour, establishing universal links between local
interactions and global phenomena. 

Among the various approaches in the complexity scientist's toolbox, information
theory has established itself as a promising framework to study distributed
computation across complex systems throughout the physical, biological, and
social
sciences~\cite{crutchfield2003regularities,lizier2012local,prokopenko2017complexity}. 
Within it, the mathematical framework of \textit{partial information
decomposition} (PID)~\cite{williams2010nonnegative,mediano2021towards}
is being rapidly consolidated as a powerful approach to understand and quantify
various aspects of distributed information processing. 
Crucially, PID provides a comprehensive taxonomy of \textit{high-order
interactions} --- statistical structures describing a group of variables that
cannot be written in terms of pairwise interdependencies.
Specifically, PID distinguishes between qualitatively different types of
high-order interactions~\cite{Timme2014}, including \textit{redundancy} ---
information
that can be extracted from multiple random variables --- and \textit{synergy}
--- information contained in a group of variables jointly but not in subsets of
them separately. 
 
A growing body of literature suggests an important role of high-order interactions supporting computation 
in complex systems. For example, alternating profiles of synergy and redundancy 
have been related to the different classes of elementary cellular 
automata~\cite{Rosas2018}, while quantities related to synergy (specifically, 
integrated information~\cite{mediano2021towards}) have been associated with 
phase transitions and distributed computation in both discrete and continuous systems~\cite{mediano2022integrated}.
In the human brain, synergy between large-scale neural populations has been
related to  cognition and consciousness, while redundancy has been associated
with sensory-motor processes~\cite{luppi2020synergistic,luppi2022synergistic}. 
In artificial neural networks, redundancy provides robustness to perturbation while synergy  
supports modality integration and flexible learning~\cite{proca2022synergistic}. 
In the context of artistic expression, the balance of redundancy and synergy has
also been related with various dimensions of musical depth and
complexity~\cite{Rosas2019,scagliarini2022quantifying}. Overall, a wealth of
empirical evidence is uncovering a spectrum of computational advantages of
high-order interactions, including links between redundancy and robustness, as
well as between synergy and flexible information integration.

These findings implicitly portray synergy as a powerful yet delicate resource,
that can provide flexibility and enhanced information-processing capabilities to
a system, while at the same time being fragile and easily broken. This picture,
however, leads to a natural question: how can this view be reconciled with the
seeming prevalence of synergy across the natural world? 
Biological systems are usually subjected to noise, i.e. stochastic fluctuations
that can perturb their dynamics in various ways. Moreover, noise is often
considered an important element in shaping dynamics and information processing
in the brain~\cite{rolls2010noisy}, allowing the encoding of different stimulus
features or improving the signal-to-noise ratio of sensory
inputs~\cite{destexhe2022noise}. Characterising the interplay between high-order
interactions and noise is an important open question, which would deepen our
understanding of how these structures may appear in physical systems.

In this paper we argue against the misconception of synergy as fragile, showing
that high-order interactions can be manifested not only in the presence of
noise, but also \textit{due to} noise. 
We illustrate our ideas with a comprehensive evaluation of the effect of noise
on the high-order structures generated by elementary cellular automata (ECA),
using the information-theoretic tools introduced in Refs.~\cite{Rosas2019} and
\cite{chechik2001group} to enable high-order analyses of relatively large
systems. Our results reveal that high-order interactions --- either of the
synergistic or redundant type --- can be enhanced by noise in the system's
dynamics, in a way that depends on the system's structure and its behaviour in
response to perturbations. 
Overall, the results in this study provide a first step towards understanding
how high-order interactions may emerge in complex distributed systems subject to
noise, including the human brain.

\section{Results}

\subsection{Entropy of cellular automata and the effect of noise}

\begin{figure}[t!]
    \centering
    \includegraphics[width=\linewidth]{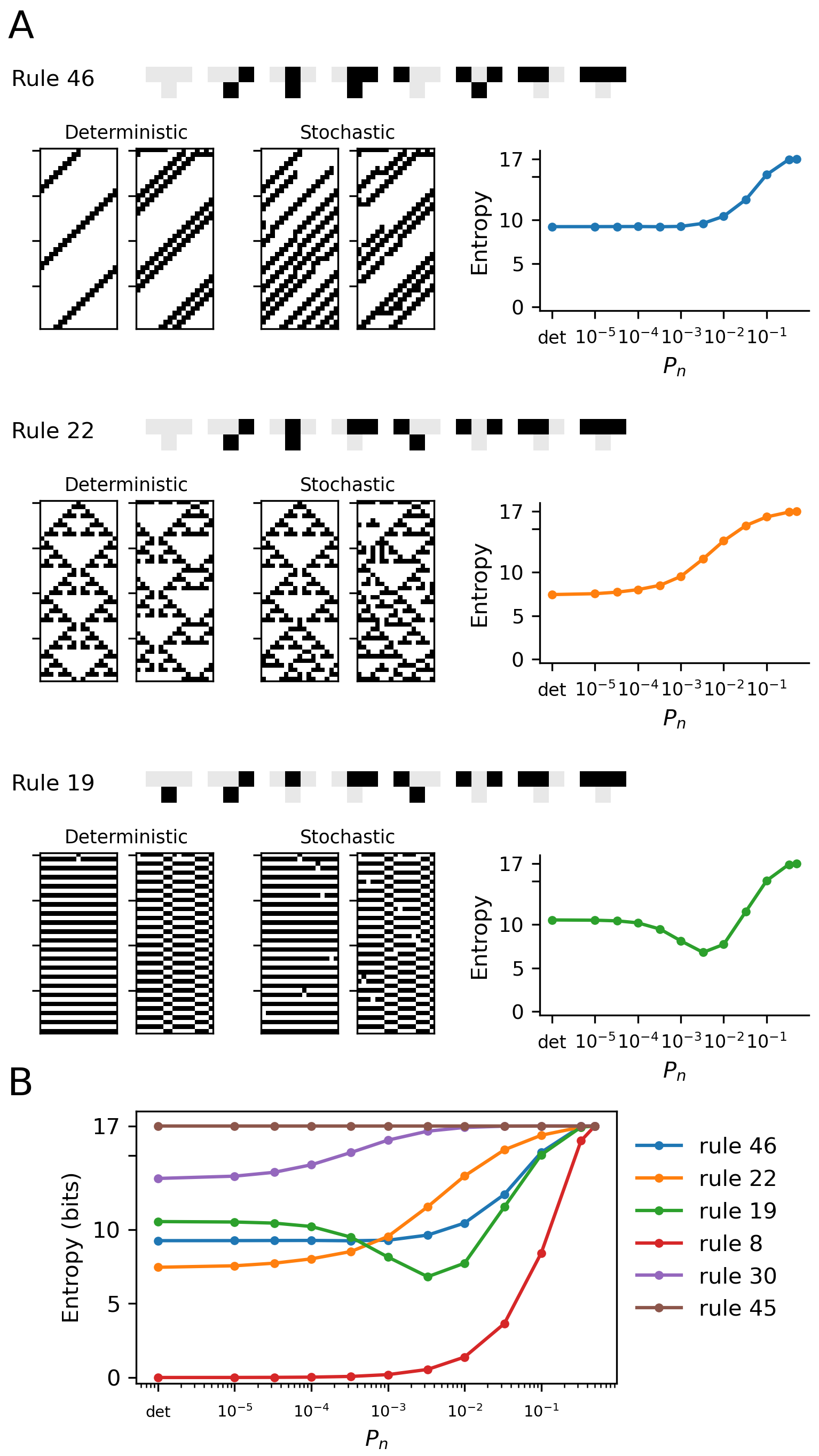}
    \caption{\textbf{Entropy of elementary cellular automata in stochastic simulations.} \textbf{(A)} Realisations of rules 46, 22, and 19 are shown in the left: 
    two trajectories without noise (deterministic), and two with probability $P_\text{n}=0.01$ of flipping an agent at each step. 
    The effect of the dynamical noise on the Shannon entropy of the resulting patterns is shown on the right. 
    \textbf{(B)} Relationship between dynamical noise and Shannon entropy for other rules discussed in the main text.}
    \label{fig:fig1_entropyNoise}
\end{figure}

The first step in our analysis was to investigate how noise impacts the diversity of the patterns resulting from the dynamics of elementary cellular automata (ECA). For this purpose, we studied the evolution of different ECA rules over a circular grid of 17 agents (or cells). 
We simulated the trajectories of all ECA rules for 800 steps considering all possible initial configurations, which allowed us to fully estimate the non-stationary joint probability distribution of agent patterns at each timestep after convergence (see \textit{Methods}). Then, the resulting patterns were characterised in terms of their joint Shannon entropy, which quantifies the diversity of configurations explored by the system. 
Simulations were carried out under different levels of dynamical noise, which was characterised in terms of the probability $P_\text{n}$ of flipping the resulting state of a given agent disobeying the corresponding ECA rule --- with $P_\text{n}=0$ being no noise, and $P_\text{n}=1/2$ being maximal noise resulting on purely random dynamics. 
It is important to highlight that the noise implemented in these analyses is \textit{dynamical noise}, i.e. perturbations to the transitions as the system's trajectory unfolds, in contrast to \textit{observational noise}, i.e. the random flipping of values of agents when observing the system but without altering its trajectory. Observational noise has a much simpler (and less interesting) effect on entropy: it simply induces a monotonic increase of it for all rules (not shown).

Results show that while adding dynamical noise often resulted in an increase of entropy, a number of ECA rules displayed a remarkable decrease of entropy for intermediate noise levels. 
Hence, maximal noise always yielded --- as expected --- patterns of maximal entropy, but the path from deterministic to fully random dynamics was found not to always be monotonous. 
To illustrate this, Figure~\subref{fig:fig1_entropyNoise}{A} shows three examples of rules with and without dynamical noise, together with the entropy measured at different values of $P_\text{n}$. This finding implies that, in some scenarios, adding noise results in a somewhat paradoxical reduction of the diversity of observed patterns. We refer to this as a `biphasic' response to noise, where a small amount of noise first creates structure, but further noise destroys it --- which we tentatively interpret in terms of a loss of metastable configurations (see Discussion).

Additionally, our results also revealed that different ECA rules display different degrees of robustness against dynamical noise, exhibiting changes in their entropy at very different noise levels. 
For example, as illustrated by Figure \subref{fig:fig1_entropyNoise}{B}, 
some rules (e.g. rule 30) are affected by very low noise levels, while others are affected only when the noise level is high (e.g. rules 46 and 8).

\vspace{-0.3cm}
\subsection{Dominance of high-order interactions can be switched by noise}
\vspace{-0.3cm}

The second step in our analysis was to investigate how dynamical noise can affect different types of high-order interactions between the agents that are generate as a result of the rules. To do this, we calculated the O-information and S-information (denoted by $\Omega$ and $\Sigma$, respectively)~\cite{Rosas2019} on the patterns exhibited by each of the ECA rules after 800 steps, as described in \textit{Methods}. These two measures provide complementary accounts of the high-order structure of a system: $\Sigma$ measures the total strength of the interdependencies in the system, while $\Omega$ measures whether these interdependencies are predominantly synergistic or redundant. In particular, a negative value of $\Omega$ indicates predominantly synergistic interactions, while a positive value indicates predominantly redundant interactionss~\cite{Rosas2019}. 

\begin{figure}[t!]
    \centering
    \includegraphics[width=\linewidth]{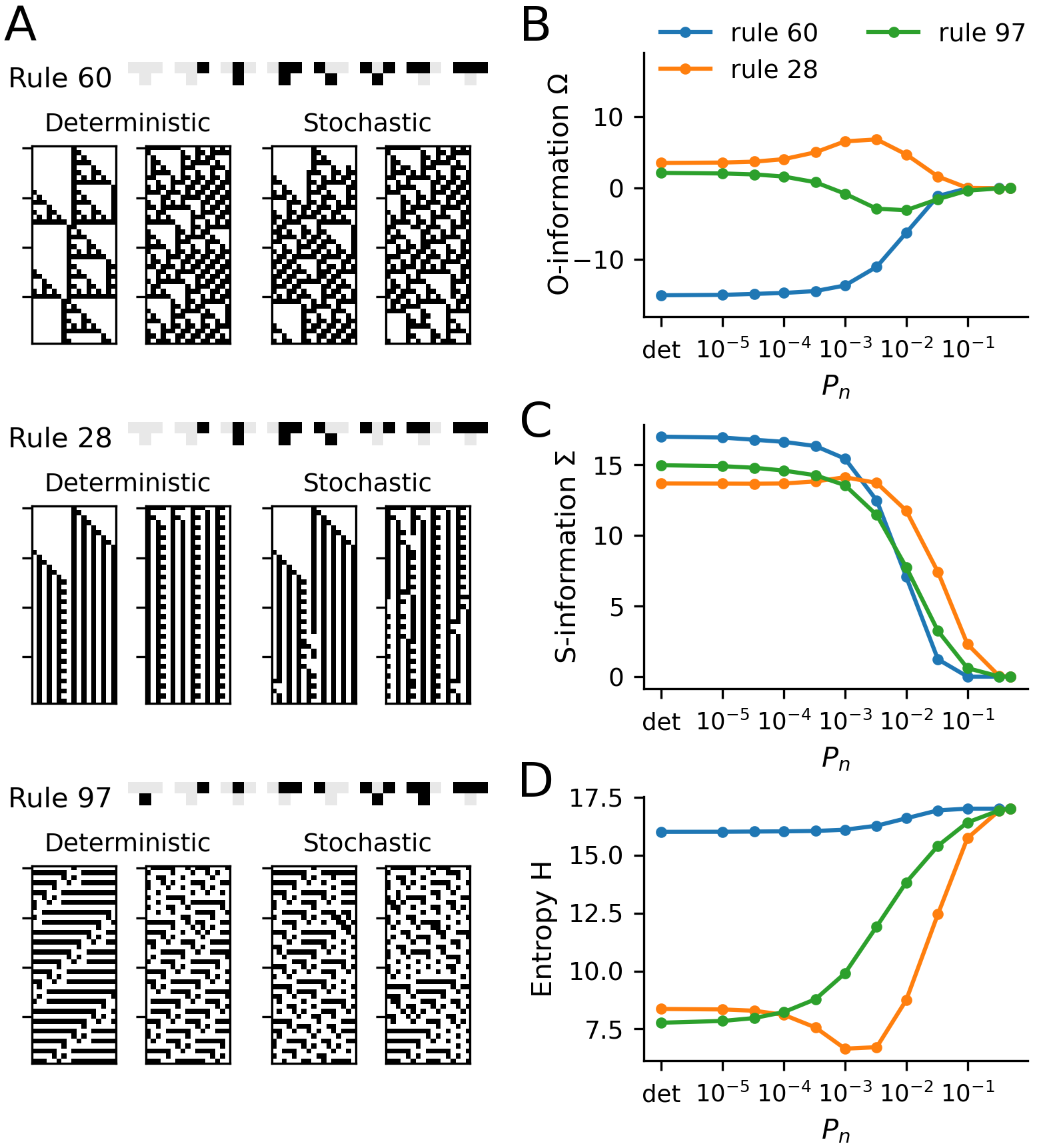}
    \caption{\textbf{High-order interdependencies in stochastic automata.} \textbf{(A)} Rules 60, 28, and 97  shown schematically with two deterministic trajectories and two stochastic trajectories ($P_\text{n}=0.01$). 
    \textbf{(B-D)} Effect of noise on the high-order structure induced by various rules as revealed by information-theoretic measures, including monotonic and biphasic behaviours.}
    \label{fig:fig2_Oinfo}
\end{figure}

Our analyses confirmed that in most ECAs, as one would expect, noise monotonically disrupts high-order structures, in the sense that both $\Omega$ and $\Sigma$ monotonically decrease in absolute value. Interestingly, however, some rules exhibit an increase in the absolute magnitude of their O-information for intermediate levels of noise. This increase can represent either a strengthening of the dominant tendency (e.g. a redundancy-dominated rule becoming even more dominanted by redundancy), or a change in the dominant tendency (e.g. a redundancy-dominated rule becoming synergy-dominated). Some examples of these phenomena are shown in Figure~\ref{fig:fig2_Oinfo}:
\begin{itemize}
    \item Rule 60
    exhibits a monotonically decreasing $\Omega$ and $\Sigma$, which is a signature of a progressive destruction of its highly synergistic structure (previously described in Ref.~\cite{Rosas2018}). 
    \item In contrast, Rule~28 is redundancy-dominated and exhibits an increase of its $\Omega$, indicating that noise makes the dominance of redundancy stronger. 
    \item Finally, Rule~97 is redundancy-dominated but its $\Omega$ becomes negative for certain values of $P_\text{n}$, indicating that for intermediate noise levels it switches to synergy-dominated. 
\end{itemize}

Additionally, for some rules (e.g. rule 28) we also found a biphasic change of
S-information with intermediate levels of noise, showing an increase before
decaying to zero (Figure~\subref{fig:fig2_Oinfo}{C}). This confirms that noise
can effectively create high-order interdependencies --- which in this case are
predominantly redundant (Figure~\subref{fig:fig2_Oinfo}{B}). In contrast, rule
97 does not show a transient increase in S-information due to noise, despite
having a biphasic behaviour of O-information
(Figure~\subref{fig:fig2_Oinfo}{B}). Correspondingly, the change in sign of
O-information may be interpreted as due to a destruction of redundant
interactions rather than the creation of synergistic ones.

\subsection{Relationship between information measures and susceptibility to noise}

Next, we sought to determine how these different information-theoretic quantities are related to each other and to the effect noise has on them. For this analysis, we excluded certain rules\footnote{Specifically, rules 0, 8, 32, 40, 128, 136, 160, 45, 75, 105, 150, 15, 51, 154, 170, and 204.} that exhibited zero O-information and S-information at all noise levels. For the remaining 77 rules, we explored relationships between entropy, O-information, and S-information, and whether the effect of noise was monotonic or biphasic. 
After noticing that different rules are affected by very different levels of noise (some start being affected at $P_\text{n}=10^{-4}$ while others only at $P_\text{n}=10^{-2}$, see Figure~\subref{fig:fig2_Oinfo}{C}), we also evaluated the ``half-noise'' level for each measure --- defined as the noise level that produces half of the total observed change --- and its relationships with information-theoretic quantities. 

\begin{figure}[t!]
    \centering
    \includegraphics[width=\linewidth]{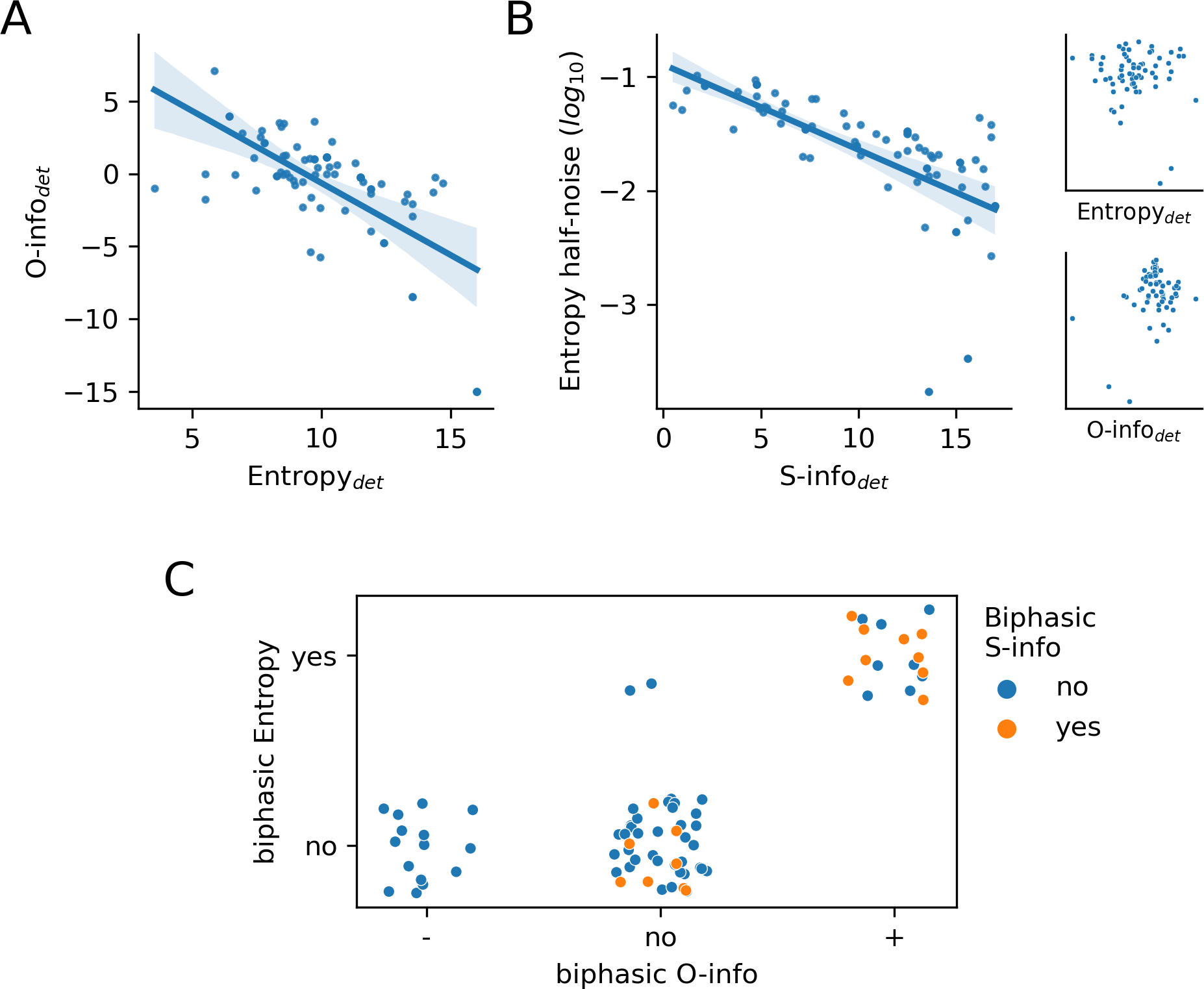}
    \caption{\textbf{Correlations between information measures in stochastic automata.} \textbf{(A)} O-information plotted against Entropy for each rule, in the deterministic setting after 800 steps of rule evolution 
    (Spearmann's $\rho=-0.6$, $p=8\times10^{-9}$). \textbf{(B)} 
    S-information in the deterministic setting versus the entropy ``half-noise,'' defined as the value of $P_\text{n}$ that causes half of the total shift in entropy ($\rho=-0.81$, $p=2\times10^{-19}$). Small plots show the absence of correlation between the entropy half-noise and entropy (top, $\rho=-0.01$, $p=0.91$) or O-information (bottom, $\rho=0.022$, $p=0.85$).  \textbf{C.} Rules classified according to the changes produced by noise on entropy (monotonic or biphasic) and O-information (negative biphasic, monotonic, or positive biphasic). The color of the dots represents the behavior of S-information (monotonic of biphasic). 
    In all plots, rules with zero or maximal entropy were excluded, as they show no effects of noise on O-information or entropy, respectively.}
    \label{fig:fig3_Correl1}
\end{figure}

Our results revealed a negative correlation between the entropy and the O-information when both are measured in the absence of noise (Figure~\subref{fig:fig3_Correl1}{A}). This means that synergy-dominated rules tend to exhibit a higher joint entropy (i.e. a larger repertoire of patterns in their final configuration), while redundancy-dominated rules 
tend to exhibit less diversity in their patterns. 
This is consistent with previous mathematical results showing that synergy tends to account for increasingly large fractions of the total entropy of a system~\cite{rosas2020operational}.
Furthermore, 
results also show a negative correlation between the half-noise for entropy change and the S-information (Figure~\subref{fig:fig3_Correl1}{B}), which implies that rules with a large amount of interdependencies are more sensitive to noise --- i.e. they are affected by smaller values of $P_\text{n}$. Interestingly, neither entropy nor O-information showed a correlation with the entropy half-noise (Figure~\subref{fig:fig3_Correl1}{B}).

Finally, results also showed that all the rules that exhibit an increase of
$\Omega$ with noise (i.e. that become more redundancy-dominated with noise) also
exhibit a biphasic change in entropy (i.e. they become less entropic with
intermediate levels of noise), as shown in Figure~\subref{fig:fig3_Correl1}{C}.
This is consistent with the result shown in Figure~\subref{fig:fig3_Correl1}{A},
where less entropy is associated with redundancy. Conversely, and with the
exception of rules 72 and 104, almost all rules that show biphasic behaviour in
entropy exhibit a positive change in O-information with noise, i.e. they become
more redundancy-dominated with noise. Finally, we found that all the rules
showing a negative drop in O-information (i.e. that transiently become more
synergistic due to noise), do not show a decrease of entropy or an increase of
S-information with noise. Overall, in agreement with our findings in the
previous section, this suggests that the increase of the predominance of synergy
caused by noise in some rules may not be accompanied by an increase of structure
--- and thus it may be due to a selective decrease of redundant interactions
instead of the creation of synergistic ones.

\subsection{Characteristic time and statistical interdependencies}

\begin{figure}[t!]
    \centering
    \includegraphics[width=\linewidth]{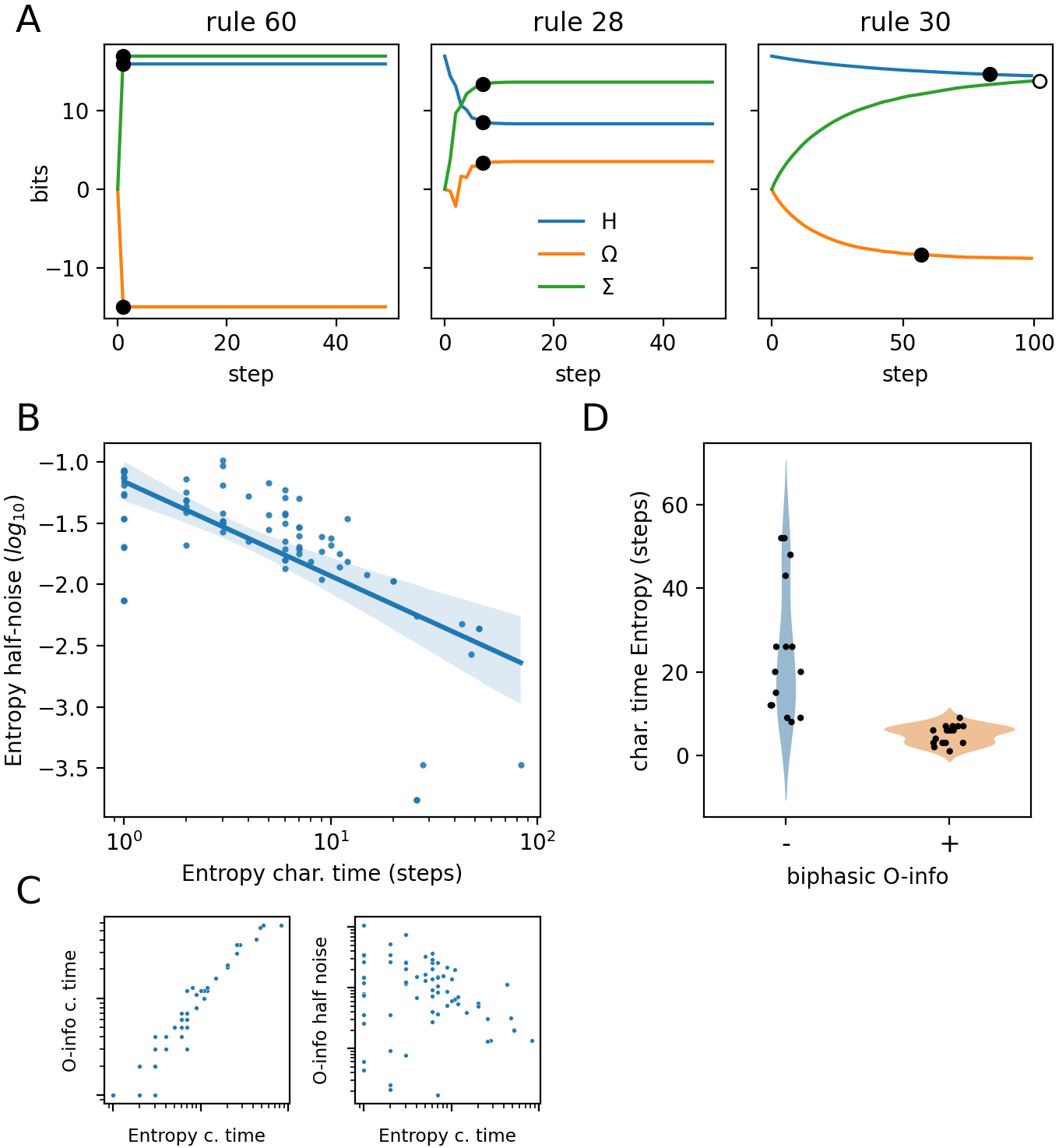}
    \caption{\textbf{Time evolution of information measures. (A)} Evolution of Entropy (H), O-information ($\Omega$)  and S-information ($\Sigma$) 
    in rules 60, 28, and 30 in the absence of noise. Black dots mark the timestep in which the measure is less than 0.5 bits away from its steady-state value, 
    which we refer to as ``characteristic time.'' For rule 30, the characteristic time for $\Sigma$ is 108 (represented as an open circle). 
    \textbf{(B)} 
    Characteristic time and logarithm of half-noise for entropy are negatively correlated (Spearmann's $\rho=-0.67$, $p=3.5\times10^{-11}$). \textbf{(C)} Left: positive correlation between the characteristic times of entropy and O-information. Right: absence of correlation between characteristic time of entropy and the half-noise for O-information. \textbf{(D)} Comparison of characteristic time for entropy for rules with either a negative or a positive biphasic O-information.}
    \label{fig:fig4_TimeCourse}
\end{figure}

We then sought to investigate how the convergence speed of various rules is related to the structure of the resulting statistical interdependencies. 
After starting from random initial conditions, some rules establish their steady states in just one or two steps (e.g. rule 60), while other rules take tens (e.g. rule 28) or hundreds (e.g. rule 110) of steps to reach stability (Figure~\subref{fig:fig4_TimeCourse}{A}). 
To quantify these differences, we defined the ``characteristic time'' of each rule as the number of steps needed to reach a difference of less than 0.5 bits (or less than 2 standard deviations for periodic rules) from the steady state value (or mean) of their joint entropy. 

We found that the characteristic time for both entropy and O-information have a negative correlation with the half-noise for entropy, as shown in Figure~\subref{fig:fig4_TimeCourse}{B}. This implies that slower rules (i.e. the ones with a larger characteristic time) have a higher susceptibility to noise, and hence less noise is needed to increase their entropy. In addition, characteristic times for Entropy and O-information are highly correlated (Figure~\subref{fig:fig4_TimeCourse}{C}, left), while there is a weak relationship between the characteristic times and and the half-noise for O-information (Figure~\subref{fig:fig4_TimeCourse}{C}, right). Finally, the rules with negative biphasic effect of noise on O-information (made predominantly synergistic with noise) are significantly slower than those with a positive biphasic effect (Figure~\subref{fig:fig4_TimeCourse}{D}). 

Overall, these results show that the rules that become predominantly redundant with noise are rules that establish their interdependencies and entropy in few steps. Taking into account the results in Figure~\subref{fig:fig4_TimeCourse}{B}, the rules that become redundant with noise are, at the same time, less susceptible to it.

\subsection{Relating high-order interdependencies to local interactions within the rules}

\begin{figure}[t!]
    \centering
    \includegraphics[width=\linewidth]{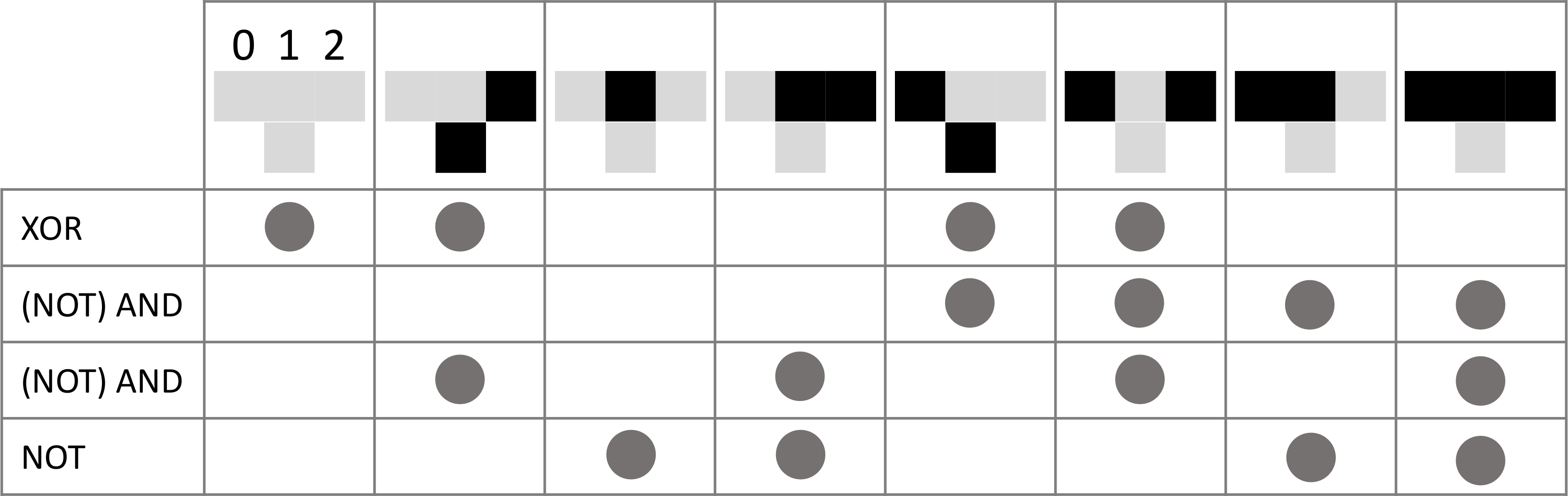}
    \caption{\textbf{Logical operations found in a rule.} Rule 18 performs the following rules: a XOR operation over the bits 0 and 2 when the bit 1 is off, a not-AND (NOR) operation over bits 1 and 2 when bit 0 is on, a NOR operation over bits 0 and 1 when bit 2 is on, and a NOT operation over bit 1. 
    }
    \label{fig:fig5_Operations}
\end{figure}

Finally, we sought to investigate how statistical interdependencies may arise
from the local interactions determined by the ECA rules. To this end, we
characterised the local interactions of each rule in two ways: 
First, we calculated the Redundancy-Synergy index (RSI, see \textit{Methods})
over each rule's three inputs bits and the output bit. The RSI quantifies the
balance between synergy and redundancy in a directed --- i.e. from inputs to
output --- manner, which is ideal to study the local effects of three
neighbouring agents determining one single output agent in the next
timepoint.\footnote{In contrast, our usage of the O-information in previous
  sections characterises the balance between redundancy and synergy over the
whole set of agents at a single timepoint in a symmetric manner.}
Second, the dependency between the inputs and the outputs was decomposed into
constituent logical operations. Specifically, for each rule we calculated the
number of binary operations of the \texttt{XOR} and \texttt{AND}/\texttt{OR}
types, as well as unary operations of the \texttt{IF}/\texttt{NOT} type. An
example is shown in Figure~\ref{fig:fig5_Operations}, which presents operations
of the three types. While most rules exhibit only one or two types, a few rules
(6, 18, 33, 40, 72, 130, 132) show the three of them. 

\begin{table*}[t]
    \centering    
\begin{tabular}{l|c|c|c||c}
Measure&\texttt{XOR}&\texttt{OR}/\texttt{AND}&\texttt{IF}/\texttt{NOT}&RSI\\
\hline
\hline
Entropy&$\times$ (p=0.76)&$\times$ (p=0.06)&$\times$ (p=0.53)&$\times$ (p=0.49)\\
O-information&\textbf{-0.39} (p=3.7e-4)&$\times$ (p=0.17)&$\times$ (p=0.29)& \textbf{-0.34} (p=0.003)\\
S-information&$\times$ (p=0.016)&$\times$ (p=0.29)& \textbf{-0.49} (p=4.1e-6)& \textbf{0.41} (p=1.7e-4)\\
\\
Entropy half-noise& \textbf{-0.46} (p=2.2e-5)&$\times$ (p=0.44)& \textbf{0.4} (p=2.7e-4)& \textbf{-0.48} (p=9.2e-6)\\
O-info half-noise&$\times$ (p=0.12)&$\times$ (p=0.63)&$\times$ (p=0.15)&$\times$ (p=0.53)\\
S-info half-noise&\textbf{-0.4} (p=2.8e-4)&$\times$ (p=0.64)&$\times$ (p=0.2)& \textbf{-0.29} (p=0.009)\\
\\
Entropy char. time&0.47 (p=1.5e-5)&$\times$ (p=0.13)&-0.53 (p=4.1e-7)&0.50 (p=2.1e-6)\\
O-info char. time&0.45 (p=3.5e-5)&$\times$ (p=0.2)&-0.48 (p=6.1e-6)&0.49 (p=5.3e-6)\\
S-info char. time&0.41 (p=1.6e-4)&$\times$ (p=0.089)&-0.53 (p=4.6e-7)&0.49 (p=5.4e-6)
\end{tabular}
    
\caption{\textbf{Linear correlation coefficients between information-based measures and local properties of rules.} \texttt{XOR}, \texttt{OR}/\texttt{AND}, and \texttt{IF}/\texttt{NOT} are the number of operations compose each rule, as depicted in Figure~\ref{fig:fig5_Operations}. RSI is the Redundancy-Synergy Index between the inputs and the rule output. Coefficients were not shown when the associated p-value was higher than 0.01.}
\label{tab:Table1_lincorr}
\end{table*}

We studied the relationship between RSI, the type of operations at the local level, O-information, and S-information. 
This analysis revealed several connections between local and collective patterns of the high-order interactions. In particular, the O-information was found to be negatively correlated with RSI --- which was expected as the RSI is (by construction) positive for synergistic interactions and negative for redundant ones. 
Moreover, the O-information was found to be negatively correlated with the number of \texttt{XOR} operations, meaning that as more of these operations are present, the rules are more synergistic. Additionally, local synergy dominance (i.e. a positive RSI) was found to be lead to higher values of S-information. The S-information was also found to be negatively correlated with the unary operations \texttt{IF} and \texttt{NOT}, meaning that unary operations make rules have less interdependencies in general. In contrast, the joint entropy was found not to be correlated to any of the local measures tested. These results are summarised in Table~\ref{tab:Table1_lincorr}.

We also investigated the relationship between RSI, local operations, the half-noise (resistance to perturbations), and characteristic time of the O-information and S-information. Results show that the entropy half-noise was negatively correlated with RSI, as well as the number of \texttt{XOR} operations, which suggests that local synergy makes a rule more susceptible to perturbations. In contrast, the presence of unary \texttt{IF/NOT} operations increase the resistance to noise, making the rule more robust. 
Furthermore, the same operations that reduce the half-noise were found to increase the characteristic times for entropy, O-information and S-information. 
Overall, rules with more synergy in the local interactions are more susceptible to noise and have longer characteristic times, while rules with unary operations are more resistant and have shorter timescales.

To assess potential combined relationships between the local rule patterns and the global information measures, we performed further analyses using Generalized Linear Models (GLM). Results largely confirm what was found using simple linear correlations: O-information is negatively correlated with the number of \texttt{XOR}s; S-information is negatively correlated with \texttt{IF/NOT}s and positively with RSI; and the half-noise is higher when there are more \texttt{IF/NOT}s and less \texttt{XOR}s (see Table~\ref{tab:Table2_GLM1}). 

\begin{table*}[]
    \centering
    \begin{tabular}{l|c|c|c|c}
Measure&GLM p-val&\texttt{XOR}&\texttt{OR}, \texttt{AND}&\texttt{IF}, \texttt{NOT}\\
\hline
\hline
Entropy& 4.1e-4& $\times$ (p=0.412)& $\times$ (p=0.0194)&\textbf{-1.76} (p=1.2e-04)\\
O-information& 0.0017&\textbf{-1.63} (p=7.6e-4)& $\times$ (p=0.425)& $\times$ (p=0.68)\\
S-information& 1.97e-08& $\times$ (p=0.503)& $\times$ (p=0.685)&\textbf{-3.24} (p=3.28e-08)\\
\\
Entropy half-noise& 4.2e-7&\textbf{-0.236} (p=5.2e-04)& $\times$ (p=0.284)& \textbf{0.174} (p=0.0081)\\
O-info half-noise& 0.305& $\times$ (p=0.242)& $\times$ (p=0.573)& $\times$ (p=0.504)\\
S-info half-noise& 0.0015&\textbf{-0.279} (p=1.7e-04)& $\times$ (p=0.205)& $\times$ (p=0.545)

    \end{tabular}
    \caption{\textbf{Generalized Linear Model fit of information-based measures and local binary or unary operations performed by each rule.} The table lists the p-value of the global fit and the coefficients (with p-values) associated to each predictor.}
    \label{tab:Table2_GLM1}
\end{table*}

%%%%%%%%%%%%%%%%%%%%%%%%%%%%%%%%%%%%%%%%%%%%%%%%%%%%%%%%%%%%
\section{Discussion}

This paper investigates the role of noise in the emergence of high-order
statistical interdependencies in complex systems, focusing on the case of
elementary cellular automata. Our results reveal that intermediate levels of
noise not only preserve, but can even enhance the statistical structure built by
certain rules --- as evidenced by a reduction of the joint Shannon entropy of
the observed patterns and corresponding increase of the S-information. This case
of ``antifragility'' (i.e. a system developing stronger structures under
external perturbations) can be tentatively interpreted as a system with unstable
attractors or transient metastable configurations. We speculate that these
configurations may become inaccessible due to perturbations, translating into a
more restricted repertoire of patterns exhibited by the system. Such systems may
have states that can only be reached with a high degree of coordination on all
the agents, which is then disrupted by dynamical noise ---  making such
transitions no longer feasible. 

Additionally, our results show that dynamical noise is capable of altering the
balance between high-order statistical structures of a system in highly
non-trivial ways. In our simulations, we found systems where noise either
enhanced the dominant high-order trend (redundant or synergistic), or changed
the dominant trend from one to another. By comparing the effect of noise on the
overall strength of high-order interdependencies (S-information) and their
dominant tendency (O-information), we observed that when structure is created by
intermediate levels of noise (i.e. a biphasic S-information profile takes place)
then it is of the redundant type (i.e. a biphasic O-information profile in the
positive direction is observed), and not synergistic. Future work may
investigate other systems to confirm if synergistic structures can be enhanced
via dynamical noise or not.

Additionally, our results reveal that systems with patterns that have more
statistical structure (as evidenced by higher S-information) are affected by
lower levels of noise. In other words, a high amount of interdependencies ---
regardless of whether they are of the synergistic or redundant type --- makes a
rule more susceptible to the either constructive or destructive effects of
noise. An explanation for this phenomenon arises from the analysis of the
temporal dynamics of the informational measures: rules with higher S-information
need in general more steps to establish its informational structure at the
beginning of the simulation. Thus, it is reasonable to expect that it may also
take longer to rebuild it when the structure is disrupted by noise.

Finally, we found correspondences between the global high-order structures and
local operations within the rules themselves. Our results showed that rules with
a synergy-dominated input-output relationship (i.e. positive RSI) tend to
generate more synergy-dominated global patterns (i.e. negative O-information),
are typically more susceptible to noise, and require a longer time to settle. On
the contrary, rules with redundancy-dominated input-output relationships or
consisting of more unary operations tend to exhibit more robustness to noise and
settle faster. These findings provide some clues related to why it is easier for
dynamical noise to enhance redundant, rather than synergistic structures.

Overall, the presented results provide a first step towards a deeper
understanding of the role of noise in the generation of high-order statistical
interdependencies, which may lead to an explanation of their pervasiveness in
the natural world. We hope these findings may inspire future efforts to
investigate similar structure-enhancing effects of noise in other complex
systems --- such as coupled oscillators, ecological models, and many others.

\vspace{-.4cm}
%%%%%%%%%%%%%%%%%%%%%%%%%%%%%%%%%%%%%%%%%%%%%%%%%%%%%%%%%%%%
%%%%%%%%%%%%%%%%%%%%%%%%%%%%%%%%%%%%%%%%%%%%%%%%%%%%%%%%%%%%
\section{Methods}
\label{methods}
\vspace{-.1cm}
%%%%%%%%%%%%%%%%%%%%%%%%%%%%%%%%%%%%%%%%%%%%%%%%%%%%%%%%%%%%
\subsection{Cellular automata simulations}
\label{sec:methods_eca}
\vspace{-.1cm}

Each elementary cellular automata rule was run for 800 steps, the number of steps chosen after checking for convergence of results on most rules. 
Noise was introduced via a probability that governed how likely an agent would disobey the rule, i.e. do the opposite that the rule dictated. A probability of 0.5 is equivalent to agents being completely random. 
For the deterministic simulations, simulations were ran for the whole set of $2^{17}$ possible initial conditions; this number was increased to 2\textsuperscript{21} for simulations considering noise, as each initial condition was repeated 8 times to allow a more thorough sampling of the possible trajectories. 
For some selected rules (those with longer characteristic times) the results were confirmed with longer simulation of 1500 steps, without finding noticeable differences. 
All simulations were carried out using custom-made code in Python. 

Among the possible 256 possible elementary rules (which involve the input of three agents), we ignored those that are equivalent up to colour inversion or mirroring transformation. Thus, only the 93 (non-equivalent) rules were analysed.\footnote{Specifically, rules 1, 2, 3, 4, 5, 6, 7, 8, 9, 10, 11, 12, 13, 14, 15, 16, 18, 19, 22, 23, 24, 25, 26, 27, 28, 29, 30, 32, 33, 34, 35, 36, 37, 38, 40, 41, 42, 43, 44, 45, 46, 50, 51, 54, 56, 57, 58, 60, 62, 72, 73, 74, 75, 76, 77, 78, 86, 90, 94, 97, 102, 104, 105, 106, 108, 110, 122, 126, 128, 130, 132, 134, 136, 138, 140, 142, 146, 150, 152, 154, 156, 160, 162, 164, 168, 169, 170, 172, 178, 184, 200, 204, and 232.}

%%%%%%%%%%%%%%%%%%%%%%%%%%%%%%%%%%%%%%%%%%%%%%%%%%%%%%%%%%%%
\subsection{Calculating information-theoretic metrics}

Following Ref.~\cite{Rosas2019} and others, the total correlation (TC) and dual total correlation (DTC) can be defined as
\begin{align}
    \text{TC}(\bm X)&=\sum_{j=1}^{n}{H(X_j)}-H\left(\bm X\right),\nonumber\\
    \text{DTC}(\bm X)&=\sum_{j=1}^{n}{H\left(\bm X^{-j}\right)}-(n-1)H\left(\bm X^n \right),\nonumber
\end{align}
where we use $\bm X = (X_1,\dots,X_n)$ and $\bm X^{-j} = (X_1,\dots,X_{j-1},X_{j+1},...X_n)$ as short-hand notations, and $H(\cdot)$ is the Shannon entropy.
Then, the O-information ($\Omega$) and S-information ($\Sigma$) can then be defined as
\begin{align}
    \Omega(\bm X) &= \text{TC}(\bm X) - \text{DTC}(\bm X),\nonumber \\
    \Sigma(\bm X) &= \text{TC}(\bm X) + \text{DTC}(\bm X).\nonumber
\end{align}

Following Ref.~\cite{chechik2001group} (see also Ref.~\cite{Timme2014}), we calculated the Redundancy-Synergy index (RSI) between the three inputs of the rule $\bm S = (S_1,S_2,S_3)$ and its output $Y$ as
\begin{equation*}
    \text{RSI}(\bm S; Y) = I(\bm S; Y) - \sum_{i=1}^3 I(S_i;Y)
\end{equation*}
where $I(S_i; Y) = H(S_i) - H(S_i|Y)$ is the mutual information between variables $Y$ and $S_i$. 

When computing these quantities, the O-information and S-information were
calculated with respect to the corresponding empirical distribution obtained by
initialising the systems in a state of maximal entropy (i.e. each agent
independent and maximally random), as described in
Section~\ref{sec:methods_eca}. The redundancy-synergy index was calculated on a
maximum entropy distribution over the inputs. We used the Plugin estimator for
deterministic simulations (as in those cases the exact joint distribution could
be computed), and the NSB estimator~\cite{nemenman2001entropy} for simulations
with noise. All information-theoretic metrics were calculated using the
\texttt{ndd} package
(\href{https://pypi.org/project/ndd}{pypi.org/project/ndd}).

%%%%%%%%%%%%%%%%%%%%%%%%%%%%%%%%%%%%%%%%%%%%%%%%%%%%%%%%%%%%
%%%%%%%%%%%%%%%%%%%%%%%%%%%%%%%%%%%%%%%%%%%%%%%%%%%%%%%%%%%%
\section*{Acknowledgements}

P.O. is funded by Agencia Nacional de Investigación y Desarrollo, ANID, Chile (grants Fondecyt 1211750 and FB0008). 
F.R. was supported by the Fellowship Programme of the Institute of Cultural and Creative Industries of the University of Kent.

%%%%%%%%%%%%%%%%%%%%%%%%%%%%%%%%%%%%%%%%%%%%%%%%%%%%%%%%%%%%
%%%%%%%%%%%%%%%%%%%%%%%%%%%%%%%%%%%%%%%%%%%%%%%%%%%%%%%%%%%%
\section*{Author contributions statement}

All the authors conceptualised the research and designed the methodology. 
P.O. ran the simulations and performed the numerical analysis. 
All the authors analysed the results, drafted, reviewed, and approved the manuscript.

%%%%%%%%%%%%%%%%%%%%%%%%%%%%%%%%%%%%%%%%%%%%%%%%%%%%%%%%%%%%
%

\end{document}